# Best Practices for a Future Open Code Policy: Experiences and Vision of the Astrophysics Source Code Library

Lior Shamir,[1] Bruce Berriman,[2] Peter Teuben,[3] Robert Nemiroff,[4] and Alice Allen[3,5]

[1] *Lawrence Technological University*
[2] *Caltech/IPAC-NExScI*
[3] *University of Maryland*
[4] *Michigan Technological University*
[5] *Astrophysics Source Code Library*

Corresponding author: Lior Shamir
lshamir@ltu.edu

**Introduction**

We are members of the Astrophysics Source Code Library's Advisory Committee and its editor-in-chief. The Astrophysics Source Code Library (ASCL, ascl.net) is a successful initiative that advocates for open research software and provides an infrastructure for registering, discovering, sharing, and citing this software. Started in 1999, the ASCL has been expanding in recent years, with an average of over 200 codes added each year, and now houses over 1,600 code entries.

Codes registered by the ASCL become discoverable not just through ASCL but also by commonly used services such as NASA's Astrophysics Data System (ADS) and Google Scholar. ASCL entries are citable and citations to them are tracked by indexers, including ADS and Web of Science (WoS), thus providing one metric for the impact of research software and accruing credit for these codes to the scientists who write them. The number of citations by ASCL ID has increased an average of 90% every year since 2013, and nearly 70 publications indexed by ADS have citations to the ASCL. Figure 1 shows the number of citations to ASCL entries by year.

The ASCL has worked to formalize source code sharing and citation practices in astrophysics (Teuben et al., 2014) and has consulted with other disciplines seeking to do the same. Our work on and experience with the ASCL, as and with software authors, and with open and closed software informs our recommendations to NASA as it contemplates establishing an open code policy.

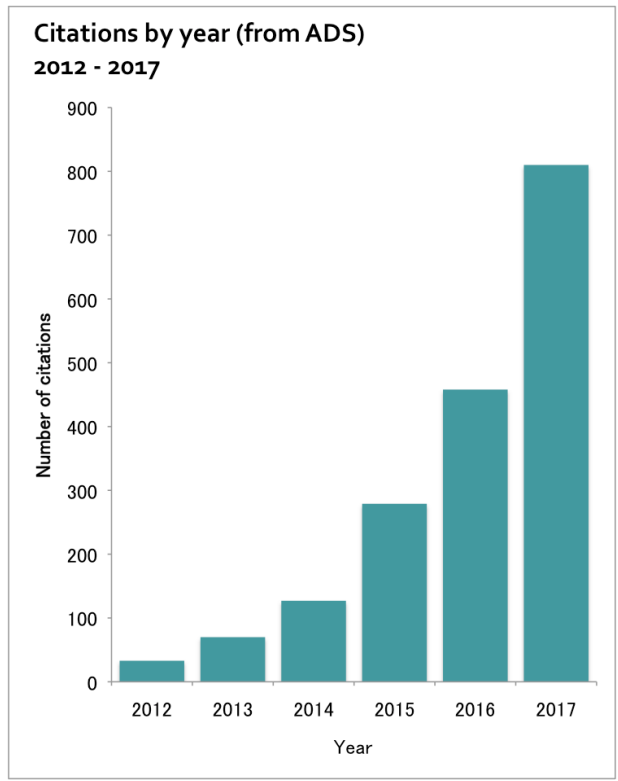

Figure 1. Citations to ASCL entries by year.

**Open Source code in astrophysics**

Due to the integration of information technology in astronomy research, computing has been becoming increasingly important, making software a pivotal part in the process of scientific conduct. Therefore, source code has become an integral part of the research, and critical for allowing the understanding, replication, and re-usability of research results. Unlike "traditional" methods of scientific communication such as peer-reviewed papers, and despite the clear need for common practices of making source code accessible to the scientific community, practices, guidelines, and requirements for making source code publicly available have not fully crystallized.

Although the requirement for making the source code open has not traditionally been a formal expectation in the field, numerous software tools have been released with Open Source Initiative[1] (OSI) licenses, making substantial impact on the field. Some notable examples include `SExtractor` (Bertin & Arnouts, 1996), `Montage` (Berriman and Good 2017), and `emcee` (Foreman-Mackey *et al.*, 2013) which has been widely used by multiple unrelated projects.

Open development, in which the source is open by design, has also been becoming more common in astrophysics; examples include the data management software stack of the Large Synoptic Survey Telescope (Jurić, 2015) and the popular `AstroPy` package (Astropy Collaboration *et al.,* 2013).

Releasing source code is important for transparency of the research, as the code is an integral part of the study and can reveal important information about the process that was carried out. Full details of the code are often very difficult to describe in a standard paper. More importantly, the availability of the code allows for replication of the results and reusability of the software, and possible discovery of unintended side effects.

Despite these important advantages, researchers are still often reluctant to release their source code. We believe the NSF Committee members, our readers, are already familiar with the reasons for this reluctance, which has been well-documented (Weiner et al. 2009, Barnes 2010, Stodden 2010, Ince et al. 2012, Katz et al. 2014, Goble et al. 2016).

Software development is a demanding task that requires a substantial amount of time, is often insufficiently funded, and is often regarded as not very important for career development decisions, such as promotion or tenure. The gap between the time required for developing a good software package and the little reward that comes with it encourages researchers to invest more time on peer-reviewed publications rather than on releasing clean documented (re)usable code. Funding software development, documentation, testing, and release would provide a financial incentive to researchers for these activities.

---

[1] https://opensource.org/

**Practices of sharing source code**

Because software is an integral part of research, the release of source code should become part of the standard scientific communication process. Currently, the primary method of scientific communications is peer-reviewed publications; the publication process can be leveraged to standardize the process of sharing source code (Shamir et al., 2013). For example, *Astronomy and Computing* recommends that authors register their research source codes with the ASCL when submitting a research paper.[2] The link between the publication process of scientific paper and the source code related to it helps to clarify requirements such as when and how source code should become available. It also makes the use of the specific source code clearer to the reader of the paper, with easier access to the specific relevant code. As peer-reviewed publications are the "building blocks" of any scientific discipline, the association between the release of the source code and the publication of the paper also standardizes the guidelines.

Since the maintenance of software after its initial goals have been accomplished is a time-consuming task yet offers little reward to the author, NASA should provide funding for maintenance of useful existing software. Standard regression tests written in the code development phase can help in software maintenance.

To provide stronger incentive for researchers to release their source code, published source codes should be indexed and become citable documents (Allen et al., 2015; this allows the software to be found on indexing engines, and also for citations for these programs to accrue to their authors. This is also consistent with the Force11 Software Citation Principles (Smith et al., 2016) and the Center for Open Science Transparency and Openness Promotion Guidelines.[3]

Funding agencies should support and encourage journals dedicated to publishing software, such as the *Journal of Open Source Software* or the *Journal of Open Research Software*. These journals allow source code and scientific software authors to receive credit equivalent to "traditional" journal publication. Clearly, these papers are also discoverable, indexed, and citable.

Research source code should be released under a licence that allows the reuse of the code, at least for non-commercial uses. Acceptable Open Source Initiative licenses, such as GPL (General Public License) and BSD (Berkeley Software Distribution) should be used to allow legal replication and use of the code with no restriction for scientific or other non-commercial purposes.

---

[2] https://www.elsevier.com/journals/astronomy-and-computing/2213-1337/guide-for-authors
[3] https://cos.io/our-services/top-guidelines/

**Recommendations**

We recommend the following:

Source code that enables research results be open for examination (released to the public) absent any truly compelling reasons, such as ITAR restrictions, that prohibit public release, upon submission of the first research paper that uses the source code to a journal. Further, we recommend that metadata about these source codes be shared with the ASCL for indexing upon submission to increase the discoverability of the software.

The version of the software used in a paper that is accepted for publication be archived in a repository before publication.

All source code developed for research be explicitly licensed with an Open Source Initiative license that permits legal replication, use, and modification of the software with no restriction for scientific or other non-commercial use.

NASA award funds explicitly for
- development of research source code, documentation, and testing procedures.
- maintenance of source code developed for research that is of continuing importance to the discipline.
- creation of new open codes that do the same tasks as closed codes. Further, we recommend undertaking a study to determine a hierarchy of codes to rewrite.

NASA require compliance with code release requirements when evaluating proposals. Further, we recommend forming a task force to 1.) develop standardized methods of reporting compliance with these requirements in new proposals where applicable, and 2.) develop instructions to referees of funding proposals to ensure these requirements are not overlooked in the evaluation process.

NASA sponsor one or more journal that publish code and software.